\documentclass[aps,prl,twocolumn,showpacs,superscriptaddress]{revtex4}
\usepackage{graphicx}
\begin{document}

\title{Temperature dependence of the paramagnetic spin excitations in BaFe$_2$As$_2$}

\author{Leland W. Harriger}
\affiliation{Department of Physics and Astronomy, The University of Tennessee, Knoxville,
Tennessee 37996-1200, USA }
\affiliation{NIST Center for Neutron Research, National Institute of
Standards and Technology, Gaithersburg, Maryland 20899, USA}
\author{Mengshu Liu}
\affiliation{Department of Physics and Astronomy, The University of Tennessee, Knoxville,
Tennessee 37996-1200, USA }
\author{Huiqian Luo}
\affiliation{Beijing National Laboratory for Condensed Matter
Physics, Institute of Physics, Chinese Academy of Sciences, Beijing
100190, China}
\author{R. A. Ewings}
\affiliation{ISIS Facility, Rutherford Appleton Laboratory, Chilton, Didcot, Oxfordshire
OX11 0QX, UK}
\author{C. D. Frost}
\affiliation{ISIS Facility, Rutherford Appleton Laboratory, Chilton, Didcot, Oxfordshire
OX11 0QX, UK}
\author{T. G. Perring}
\affiliation{ISIS Facility, Rutherford Appleton Laboratory, Chilton, Didcot, Oxfordshire
OX11 0QX, UK}
\author{Pengcheng Dai}
\email{pdai@utk.edu}
\affiliation{Department of Physics and Astronomy, The University of Tennessee, Knoxville,
Tennessee 37996-1200, USA }
\affiliation{Beijing National Laboratory for Condensed Matter
Physics, Institute of Physics, Chinese Academy of Sciences, Beijing
100190, China}

\begin{abstract}
We use inelastic neutron scattering to study temperature dependence of the paramagnetic spin excitations
in iron pnictide BaFe$_2$As$_2$ throughout the Brillouin zone.  In contrast to a conventional local moment Heisenberg system,
where paramagnetic spin excitations are expected to have
a Lorentzian function centered at zero energy transfer, the
high-energy ($\hbar\omega>100$ meV) paramagnetic spin excitations in BaFe$_2$As$_2$
exhibit spin-wave-like features up to
at least 290 K ($T= 2.1T_N$).   Furthermore, we find that the sizes of the fluctuating magnetic
moments $\left\langle m^2 \right\rangle\approx 3.6\ \mu^2_B$ per Fe
are essentially temperature independent from the AF ordered state at $0.05T_N$
to $2.1T_N$, which differs considerably from the temperature dependent fluctuating moment observed in the iron chalcogenide Fe$_{1.1}$Te [I. A. Zaliznyak {\it et al.}, Phys. Rev. Lett. {\bf 107}, 216403 (2011).].  These results suggest unconventional magnetism and strong electron correlation effects in BaFe$_2$As$_2$.
\end{abstract}

\pacs{75.30.Ds, 25.40.Fq, 75.50.Ee}

\maketitle

The elementary magnetic excitations (spin waves and paramagnetic spin excitations) in a ferromagnet or
an antiferromagnet can provide direct information about
the itinerancy of the unpaired electrons contributing to the ordered moment.  In a local moment system, spin waves are usually well-defined throughout the Brillouin zone and can be accurately described by a Heisenberg Hamiltonian
in the magnetically ordered state.  The total moment sum rule requires that the dynamical structure factor $S(q,\omega)$, when integrated over all wave vectors ($q$) and energies ($E=\hbar\omega$),
is a temperature independent constant and equals to $\left\langle m^2\right\rangle=(g\mu_B)^2S(S+1)$, where $g$ is the Land$\rm \acute{e}$ $g$ factor ($\approx 2$) and
$S$ is the spin of the system \cite{lorenzana}.  Upon increasing temperature to the paramagnetic state,
spin excitations in the low-$q$ limit can be described by a simple Lorentzian scattering function $S(q,\omega)\propto [1/(\kappa_1^2+q^2)][\Gamma/(\Gamma+\omega^2)]$,
where $\kappa_1$ is the temperature dependent inverse spin-spin correlation length 
 and $\Gamma$ is the wave vector dependent characteristic energy scale \cite{tucciarone,wicksted,jwlynn}.
At sufficiently high temperatures above the magnetic order, spin excitations should be purely paramagnetic with no spin-wave-like correlations.
Therefore, a careful investigation of the wave vector and energy dependence of spin excitations across the magnetic ordering
temperature can provide important information concerning the nature of the magnetic order and spin-spin correlations.  For example, a recent inelastic neutron scattering study of spin excitations in
one of the parent compounds of iron-based superconductors,
the iron chalcogenide Fe$_{1.1}$Te which has a bicollinear antiferromagnetic (AF) structure and N$\rm \acute{e}$el temperature of $T_N=67$ K
\cite{Hsu,Fang,Bao,Li,lipscombe,Johnston,dai}, reveals that the effective spin per Fe changes from $S\approx 1$ in the AF state to $S\approx 3/2$ in the paramagnetic state, thus providing evidence that Fe$_{1.1}$Te is not
a conventional Heisenberg antiferromagnet but a nontrivial local moment system coupled with itinerant electrons \cite{igor}.

Since antiferromagnetism may be responsible for electron pairing and superconductivity in iron-based superconductors \cite{scalapino,Hirschfeld}, it is important to determine if the observed anomalous
spin excitation behavior in iron telluride Fe$_{1.1}$Te is a general phenomenon in the parent compounds of
iron-based superconductors.  For this purpose, we study the spin excitations of another parent compound of iron-based superconductors, the
iron pnictide BaFe$_2$As$_2$ which has a collinear AF structure with
$T_N\approx 138$ K \cite{Kamihara,cruz,rotter,qhuang}, over a wide temperature range ($0.05T_N\leq T\leq 2.1T_N$).  In the low-temperature orthorhombic phase ($T=0.05T_N=7$ K),
 previous inelastic neutron scattering experiments found that a Heisenberg Hamiltonian with
 highly anisotropic effective magnetic exchange couplings and damping along the orthorhombic $a$ and $b$ axes directions  can describe the observed spin-wave spectra \cite{lharriger}, similar to
the spin waves in the collinear AF ordered CaFe$_2$As$_2$ \cite{jzhao}.
However, similar measurements on iron pnictide SrFe$_2$As$_2$ suggest that spin waves can be better described
by calculations from a five-band itinerant mean-field model \cite{ewings,kaneshita}.  Therefore, it is unclear whether a localized Heisenberg Hamiltonian \cite{lharriger,jzhao} or itinerant magnetism \cite{ewings,kaneshita,diallo09} is
a more appropriate description for spin waves in pnictides.  In the high-temperature tetragonal phase, the spin-wave anisotropy of BaFe$_2$As$_2$
appears to persist at 150 K ($T=1.1T_N$) suggesting
 the presence of an electronic nematic phase \cite{lharriger,fradkin,tmchuang,jhchu,myi,kasahara,ruff}. However, these paramagnetic
 spin excitations can also be understood by considering both the localized and itinerant electrons
 using dynamic mean field theory \cite{hpark} or a biquadratic spin-spin interactions within a Heisenberg Hamiltonian without the need for electronic nematicity \cite{wysocki,ryu}.

By studying spin excitations in BaFe$_2$As$_2$ over a wide temperature range throughout the Brillouin zone in absolute units, we
can determine the temperature dependence of the paramagnetic scattering and its spectral weight.
This will reveal if itinerant electrons in BaFe$_2$As$_2$ are coupled
with local moments on warming across $T_N$
similar to that of the iron telluride Fe$_{1.1}$Te \cite{igor}. Surprisingly,
we find that the total fluctuating magnetic moments $\left\langle m^2\right\rangle\approx 3.6\ \mu_B^2$ per
Fe in BaFe$_2$As$_2$, corresponding to an effective spin $S=1/2$ per Fe \cite{msliu}, are essentially unchanged on warming from 7 K at $T=0.05T_N$
to room temperature at
$2.1T_N$, much different from that of Fe$_{1.1}$Te \cite{igor}.
In addition, while paramagnetic spin excitations at small wave vectors near the AF zone center follow
a simple Lorentzian scattering function as expected \cite{tucciarone}, they change only slightly 
from the low-temperature spin waves 
for wave vectors near the zone boundary up to room temperature.  This is different from the expectation of a local moment Heisenberg system, and indicate a strong electron correlation effect in BaFe$_2$As$_2$.

\begin{figure}
\begin{center}
\includegraphics[width=0.75\linewidth]{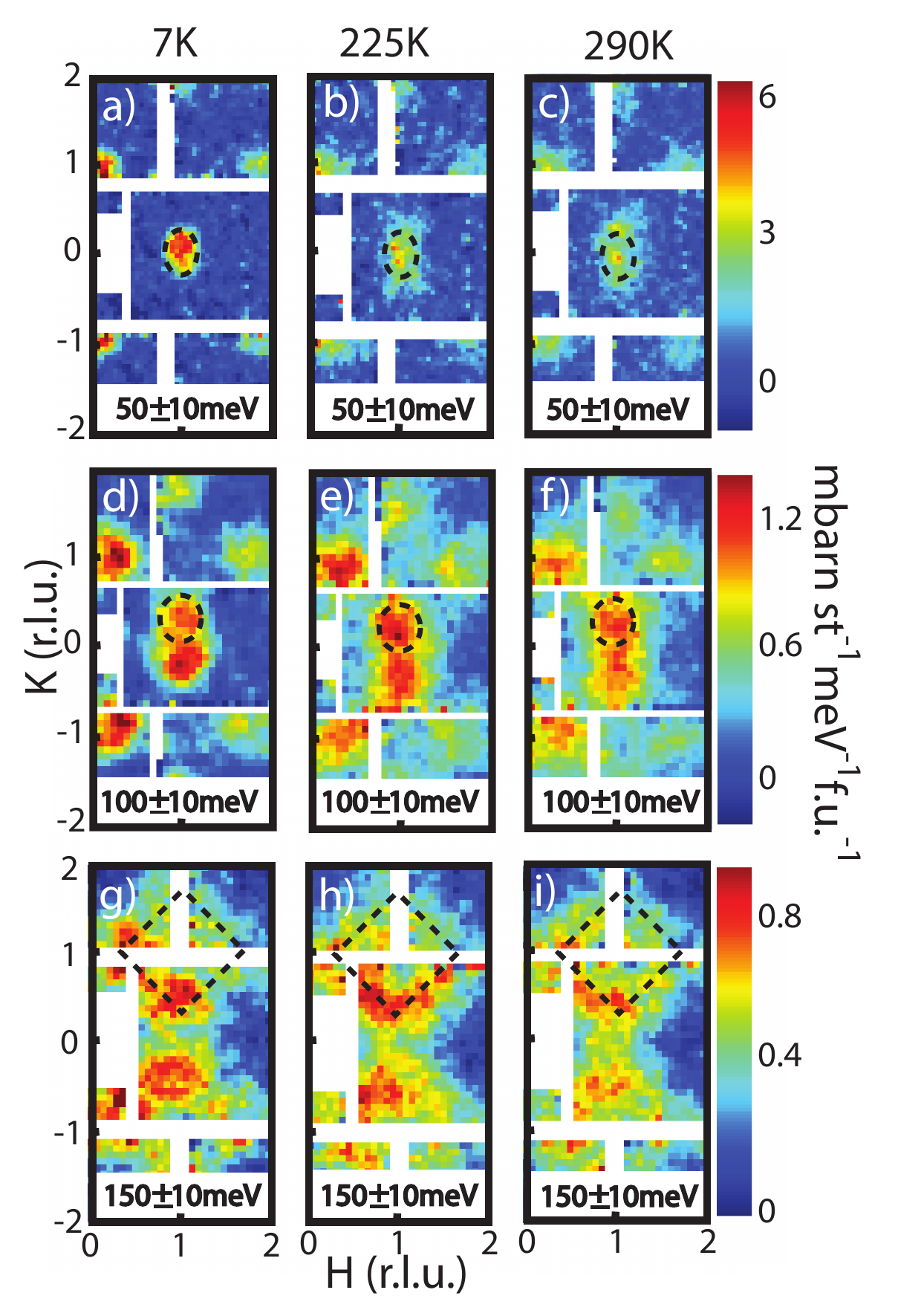}
\end{center}\caption{
(Color online) Panels a)-c) compare the $E=50$meV magnetic scattering deep inside the ordered state (7 K) to scattering in the paramagnetic phase for temperatures (225 K and 290 K) well away from the $T_{N} = 138$ K phase transition. Panels d)-f) and g)-i) are a similar comparison for energy transfers of 100 meV and 150 meV, respectively. The dotted ellipses and boxes are guides to the eye to more easily facilitate comparison. Data in panels a)-f) and g)-i) were collected using $E_i=250$ meV and 450 meV respectively. All data was background subtracted using the average intensity from the region $1.8<H<2.2$, $-0.2<K<0.2$ r.l.u. as the background point. Data in the region $H<0$ was folded into the equivalent $H>0$ positions in order to improve statistics.
}
\label{Fig:fig1}
\end{figure}

We have used the MAPS time-of-flight inelastic neutron spectrometer at ISIS, Rutherford-Appleton Laboratory, UK, to determine the paramagnetic excitations of BaFe$_2$As$_2$.
For the experiment, we have used the same sample and experimental set-up as described previously \cite{lharriger}.
Below $T_N$, BaFe$_2$As$_2$ has an orthorhombic structure with $a=5.62$ and $b=5.57$ \AA\ 
and forms a collinear AF order at the ordering wave vector $Q=(1,0,1)$ \cite{qhuang}.  In the paramagnetic state, BaFe$_2$As$_2$ changes to
tetragonal structure.  Figure 1 presents an overview of the temperature evolution of the spin excitations at different energies.
The data has been normalized to a vanadium standard and plotted in absolute units of mbarn sr$^{-1}$ meV$^{-1}$ f.u.$^{-1}$, without correction for the magnetic form factor, leading to a decrease in
magnetic scattering with increased $Q$.  At $E=50\pm 10$ meV, spin waves form ellipses along the transverse direction centered at
$Q_{\rm{AF}}=(1+m,n,L)$ and  $Q_{\rm{AF}}=(m,1+n,L)$, where $m,n=0,1,2,\cdots$ and $L=1,3,5,\cdots$, at 7 K (Fig. 1a) \cite{lharriger}.
Upon warming to the paramagnetic state at 225\,K ($T=1.63T_N$, Fig. 1b) and 290 K ($2.1T_N$, Fig. 1c), the signal becomes weaker, and the ellipses broader, compared to the spin wave peak seen at 7\,K, similar to the low-energy paramagnetic spin excitations seen in CaFe$_{2}$As$_{2}$ \cite{diallo}.
However, the spin waves at $E=100\pm 10$ (Figs. 1d-1f) and $150\pm 10$ meV (Figs. 1g-1i) only decrease slightly in intensity on warming, and become more diffusive at 290 K.

\begin{figure}
\begin{center}
\includegraphics[width=0.8\linewidth]{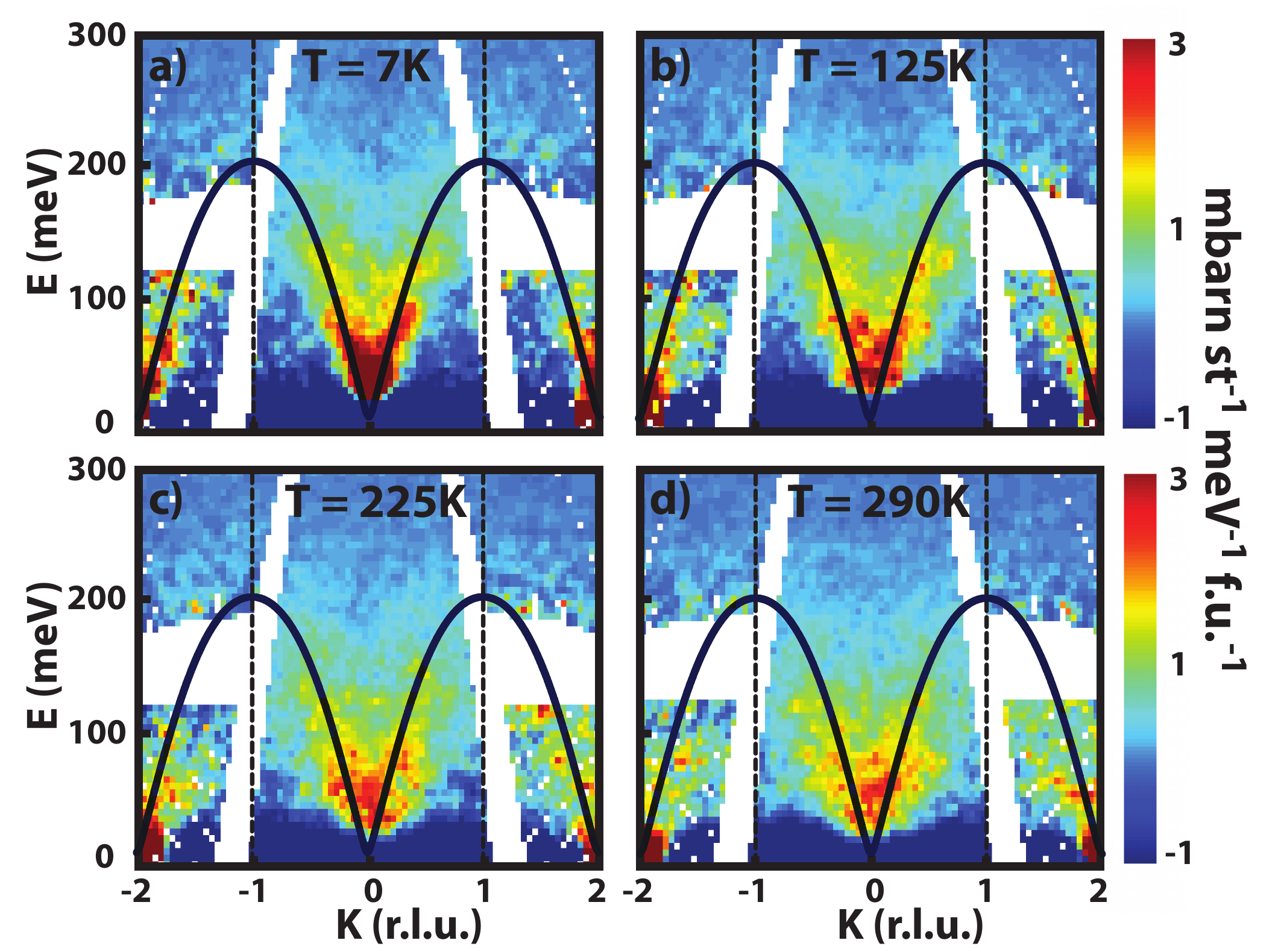}
\end{center}\caption{
(Color online) Panels a)-d) compare the Energy versus $K$ intensity slices for 7 K, 125 K, 225 K, and 290 K. All data is background subtracted and folded in an identical manner as described in the caption of Fig. \ref{Fig:fig1}. The solid line is the Heisenberg dispersion obtained using anisotropic exchange couplings $SJ_{1a} = 59.2 \pm 2.0$, $SJ_{1b} = -9.2 \pm 1.2$, $SJ_{2} = 13.6 \pm 1.0$, $SJ_{c} = 1.8 \pm 0.3$ meV determined by fitting the full cross section to the 7 K data \cite{lharriger}.
}
\label{Fig:fig2}
\end{figure}

Figures 2a-2d show the background subtracted scattering for the
$E_i = 450$ meV data projected in the wave vector (${\bf Q} = [1, K]$) and energy space
at $T=7$, 125, 225, and 290 K, respectively.  The solid lines represent the expected dispersion from the anisotropic Heisenberg Hamiltonian \cite{lharriger}.
At 7 K, three plumes of spin waves stem
from $Q_{\rm{AF}} = [1, K]$ where $K=0,\pm2$, and reach to
the zone boundary at $\sim$200 meV (Fig. 2a).  On warming to 125 (Fig. 2b), 225 (Fig. 2c), and 290 K (Fig. 2d), spin excitations become broader in
momentum space but their zone boundary energies appear to be unchanged.  For the classical insulating Heisenberg ferromagnet or antiferromagnet,
spin excitations in the paramagnetic state should be uncorrelated and
display Lorentzian-like peaks centered at $E=0$ meV at sufficiently high temperatures
\cite{tucciarone,wicksted}.  If electron correlations are important, spin excitations in the paramagnetic state should exhibit spin-wave like peaks
 in energy for wave vectors near the zone boundary  \cite{jwlynn}.
While previous work found that spin excitations near the zone boundary for energies above $E=100$ meV are indeed similar between 7 K and
150 K \cite{lharriger}, it is unclear what happens to zone boundary spin excitations
at higher temperatures.

\begin{figure}
\includegraphics[width=0.9\linewidth]{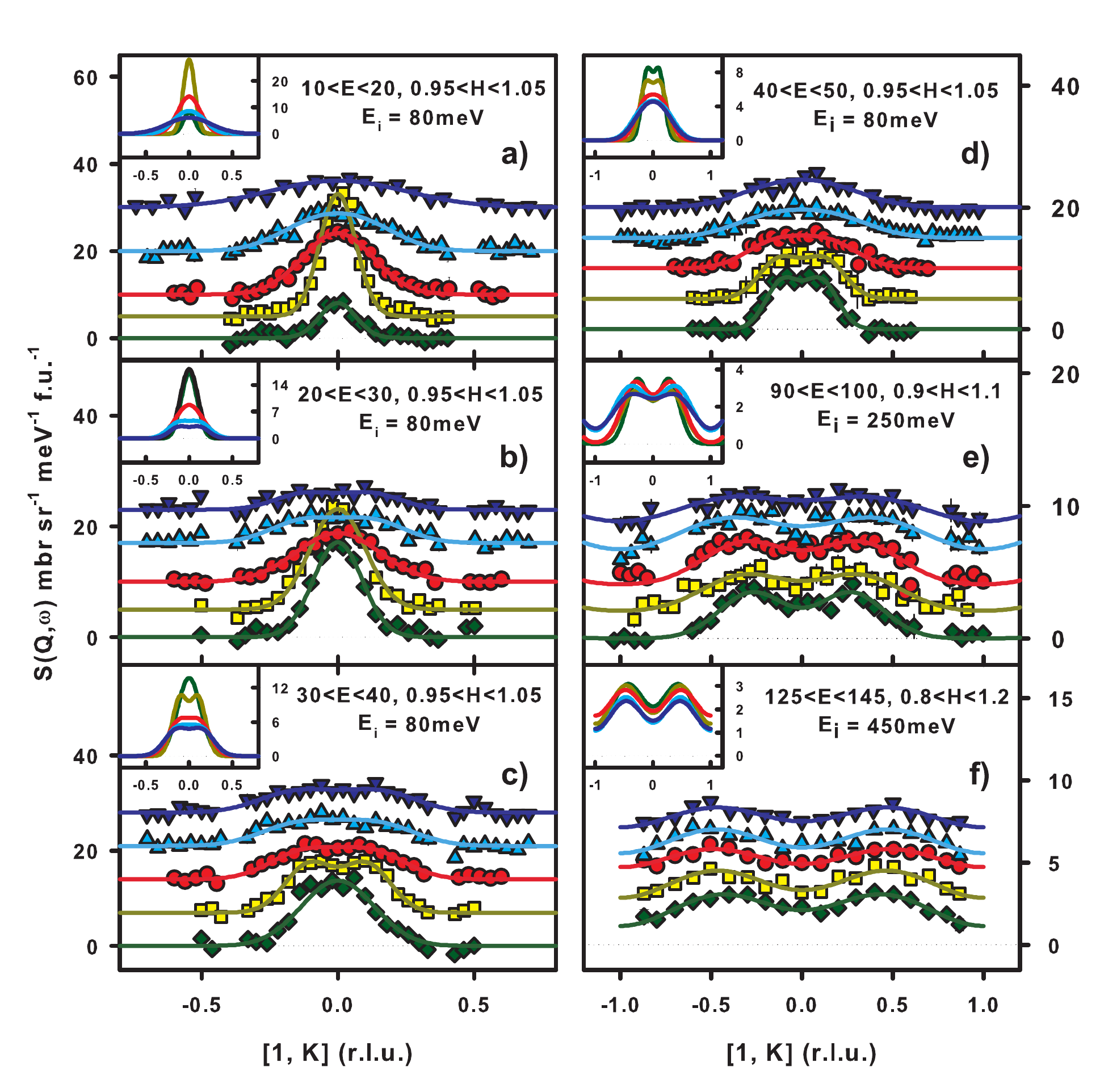}
\caption{(Color online) a-f) Temperature overplot of the evolution of the spin excitations as a function of increasing energy. The green diamonds, yellow squares, red circles, cyan upward facing triangles, and blue downward facing triangles are for the 7 K, 125 K, 150 K, 225 K, and 290 K data respectively. The data has been artificially offset for clarity and empirically fit using Gaussian functions. The insets are the fits without offset.
}
\label{Fig:fig3}
\end{figure}

Figure 3 summarizes the wave vector and temperature dependence of the
spin excitations from 7 K to 290 K along the ${\bf Q}=[1,K,0]$ direction.  For each of the wave vector cuts along the $K$-direction,
the $H$-direction integration range is slightly different.  At $10\leq E\leq 20$ meV, the spin wave intensity increases on warming from 7 K to 125 K.  Upon further warming to above $T_N$, the spin excitation peak centered at $Q_{\rm{AF}}=(1,0,L)$ becomes weaker and broader with increasing temperature, and 
is very broad at 290 K.  
For spin wave energies $20\leq E\leq 30$ meV (Fig. 3b), $30\leq E\leq 40$ meV (Fig. 3c),
and $40\leq E\leq 50$ meV (Fig. 3d),
the situation is similar although spin excitations have less temperature dependence with increasing energy.
Finally, spin excitations only change marginally from 9 K to 290 K for $90\leq E\leq 100$ meV (Fig. 3e) and $125\leq E\leq 145$ meV (Fig. 3f).

\begin{figure}
\begin{center}
\includegraphics[width=1.0\linewidth]{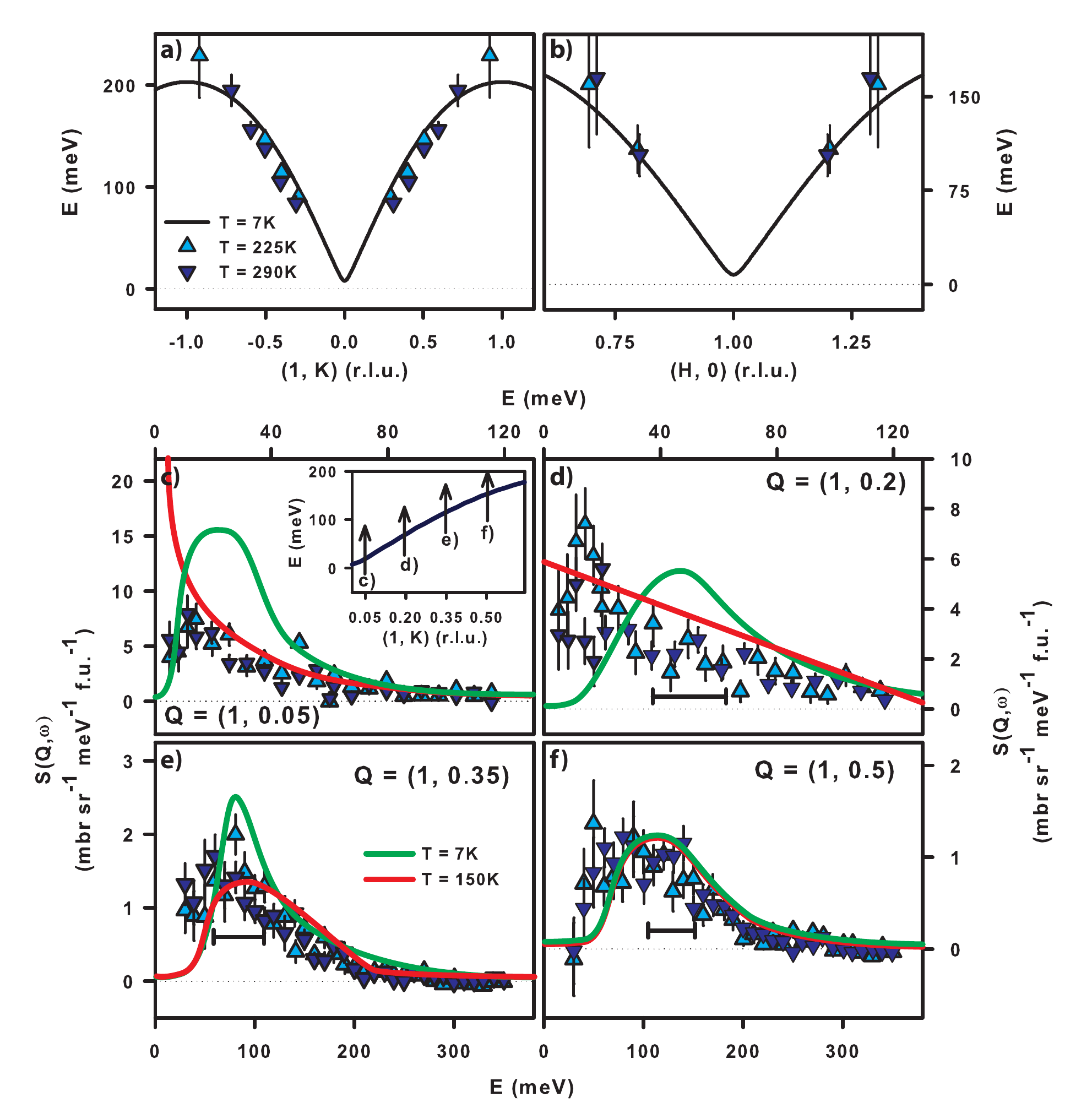}
\end{center}\caption{
(Color online) a) Dispersion along the $[1, K]$ direction as determined by energy and $Q$ cuts of the raw data. The solid line is the anisotropic Heisenberg dispersion \cite{lharriger}. b) Dispersion along the $[H, 0]$ direction built using the same method. The light blue upward facing triangular points in c)-f) are constant-$Q$ cuts at $Q = (1, 0.05)$, (1, 0.2), (1, 0.35), and (1, 0.5), respectively, at 225 K. The dark blue downward facing triangular points in c)-f) are identical constant-$Q$ cuts at 290 K. The solid green and red lines are guides to the eye describing the observed 7 K and 150 K scattering, respectively. These constant-$Q$ cuts correspond to cuts across the dispersion as depicted in the inset of panel c). The horizontal bars in d) and f) are instrumental energy resolution.
}
\label{Fig:fig4}
\end{figure}

Based on the data in Figures 2 and 3, we construct in Fig. 4a spin
excitation dispersions along the $[1,K]$ and $[H,0]$ directions
at 225 K (the upper triangles) and 290 K (the lower triangles).  Comparing the outcome
with the spin waves at 7 K (the solid lines) reveals essentially the same dispersion for spin excitations in the paramagnetic state for temperatures
up to $T=2.1T_N$ (Figs. 4a and 4b).  Figures 4c-4f show constant-$Q$ cuts of the spin excitations
along the $[1,K]$ direction throughout the Brillouin zone (see inset in Fig. 4c).
Previous measurements at 7 K and 150 K are plotted as green and red solid lines, respectively.
For wave vector near the zone center at $Q=(1,0.05)$ and $(1,0.2)$, we see that the well-defined spin wave peaks
in the AF phase become Lorentzian like in the paramagnetic state at 150 K (Fig. 4c and 4d).
On further warming to 225 and 290 K, quasielastic  intensity near $E=0$ meV
becomes weaker, consistent with the expectations for paramagnetic scattering \cite{tucciarone,wicksted,jwlynn}.
However, the low-temperature spin wave peaks
at the wave vectors $Q=(1,0.35)$ and $(1,0.5)$ near the zone boundary
are still clearly present up to 290 K, and only become
slightly broader and weaker (Figs. 4e and 4f), thus suggesting a
strong electron correlation effect in BaFe$_2$As$_2$.

\begin{figure}
\begin{center}
\includegraphics[width=0.90\linewidth]{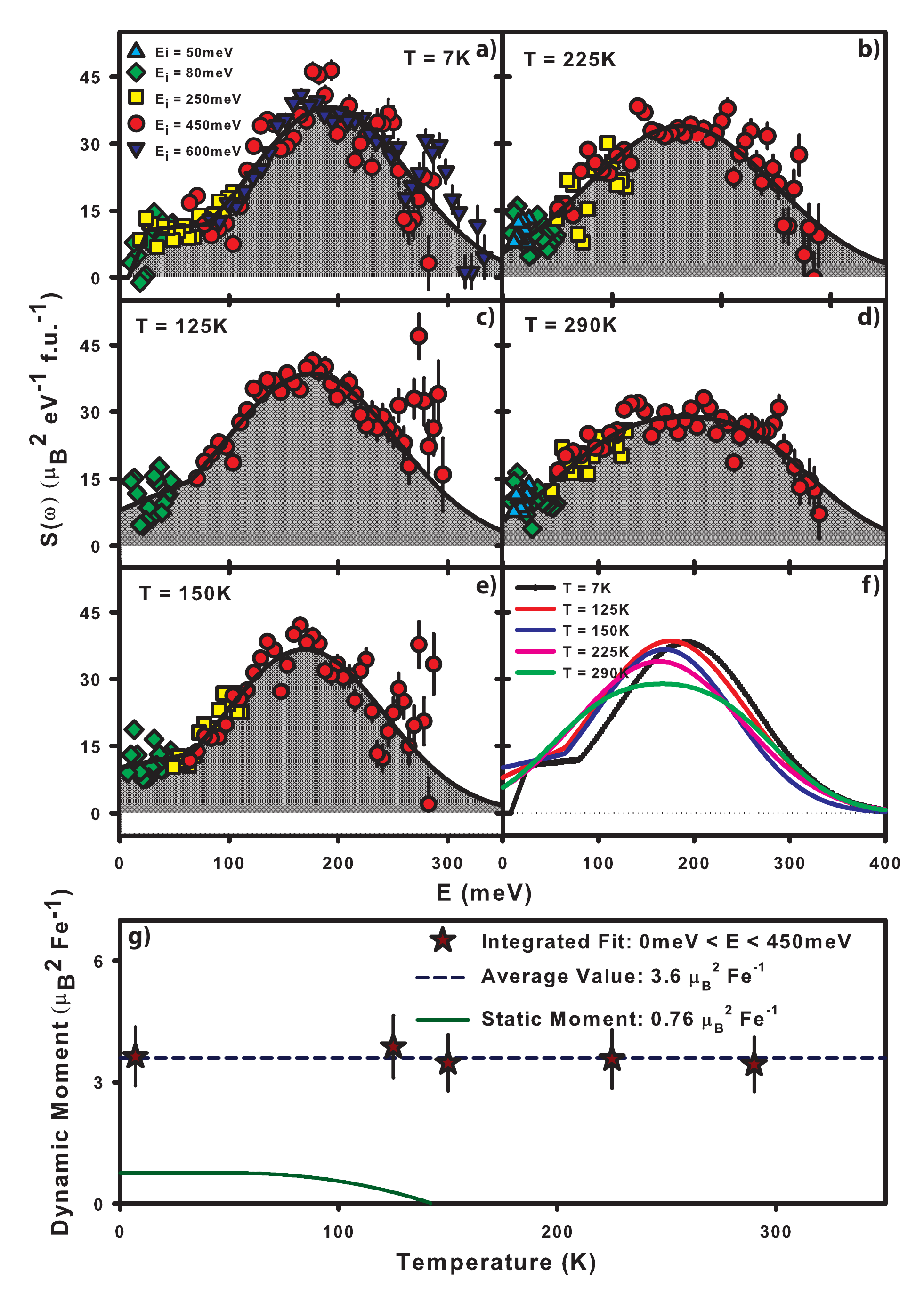}
\end{center}\caption{
(Color online) The local susceptibility plots in panels a)-e) represent the total $Q$-integrated intensity across the magnetic zone of size $0<H\leq 2$, $-1<K\leq 1$ r.l.u. In practice, it is not possible to use the actual full zone size in $H$ and $K$ because of 
gaps in the detector array and consequent limited accessibility of certain reciprocal space regions. 
Thus, for each $E_{i}$ a smaller region that either contains all of the scattering and/or has the requisite symmetry is chosen instead. These regions are then all normalized to the entire zone area as required by $\chi^{\prime\prime}(E) = \frac{\int\chi^{\prime\prime}({Q},E)d{Q}}{\int d{Q}}$. The solid black lines in panels a)-e) are empirical fits of the local susceptibility. f) An overplot of these fits to aid in a cross-comparison of the temperature dependence. g) The dynamic moment as determined by integrating the fits from the previous panels. The static moment is reproduced from \cite{qhuang}.
}
\label{Fig:fig5}
\end{figure}

Finally, we show in
Figure 5 the temperature dependence of the local dynamic susceptibility for BaFe$_2$As$_2$ \cite{msliu,clester}.  In the AF ordered state at 7 K,
there is a spin anisotropy gap below $\sim$10 meV \cite{matan} and the local susceptibility peaks at $\sim$180 meV (Fig. 5a).  On warming to 125 K just below
$T_N$, the spin anisotropy gap disappears while at higher energies the local susceptibility remains essentially unchanged (Fig. 5c).
Upon further warming to the paramagnetic state at 150 K (Fig. 5e), 225 K (Fig. 5b), and 290 K (Fig. 5d),
we see that the local dynamic susceptibility becomes slightly weaker and
broader with increasing temperature (Fig. 5f).
Figure 5g shows the temperature dependence of the ordered moment (solid line) \cite{qhuang}
and integrated local susceptibility,
which is dominated by spectral weight from spin excitations above 100 meV.
For comparison,
we note that the integrated magnetic spectral weight of Fe$_{1.1}$Te were reported to
concentrate almost entirely within 30 meV \cite{igor}.

In earlier triple-axis spectrometry studies of paramagnetic spin excitations of metallic ferromagnets such as iron and nickel, there was considerable controversy concerning whether persistent spin wave like excitations can exist in the paramagnetic state above $T_C$ \cite{wicksted,mook,lynn75,steinsvoll,endoh}.
 For BaFe$_2$Aa$_2$, we see spin-wave-like excitations above 100 meV at temperatures up to $2.1T_N$.
This is different from the usual paramagnetic scattering in a Heisenberg antiferromagnet.  
The lack of temperature dependence of the integrated
local moment, $\left\langle m^2\right\rangle\approx 3.6\ \mu_B^2$ per
Fe, suggests that the effective spin of iron in BaFe$_2$As$_2$ ($S=1/2$) is unchanged from the AF
orthorhombic phase to the paramagnetic tetragonal phase up to room temperature.
Therefore, there is no exotic entanglement of itinerant electrons with localized magnetic moments, much
different from that of the Fe$_{1.1}$Te \cite{igor}.  We also note that the size of the dynamic moment 
$\sqrt {\left\langle m^2\right\rangle}\approx 1.9\ \mu_B$ per
Fe in BaFe$_2$As$_2$ is larger than the local moment of $1.3\ \mu_B$ per Fe
determined from x-ray emission
spectroscopy \cite{gretarsson}, but similar to the local moment of $2.1\ \mu_B$ per Fe in SrFe$_2$As$_2$ obtained from 
the Fe $3s$ core level photoemission spectra measurements \cite{vilmercati}.

In summary, we have studied the temperature dependent paramagnetic spin
excitations in iron pnictide BaFe$_2$As$_2$, one of the parent compounds of iron-based superconductors.
In contrast to a conventional Heisenberg system, we find spin-wave-like paramagnetic excitations near the zone boundary for temperatures up to $2.1T_N$ with no evidence for the expected
zone boundary magnon softening.  In addition, the integrated local magnetic moment is remarkably temperature independent from the AF ordered orthorhombic phase to the paramagnetic tetragonal phase,
and corresponds to an effective 
iron spin of $S=1/2$.  This is different from
the temperature dependent spin excitations in the iron chalcogenide Fe$_{1.1}$Te.  Our results indicate a strong electron correlation effect and suggest that the entanglement of itinerant electrons with localized magnetic moments in Fe$_{1.1}$Te \cite{igor} is not fundamental to the magnetism in the parent compounds of iron-based superconductors. Furthermore, correctly modeling the pnictides requires taking into account a mixed state where correlations are important. Indeed, 
both dynamic mean field theory \cite{hpark} and biquadratic exchange \cite{wysocki,ryu} 
are approaches that pick up electron correlations and appear to provide necessary features 
 for describing the physics of these systems. 

We are grateful to Jeffrey Lynn for helpful discussions.
The work at UTK is supported by the US NSF DMR-1063866. 
Work at the IOP,CAS is supported by the MOST of China 973
programs (2012CB821400, 2011CBA00110) and NSFC-11004233.

\end{document}